\documentclass[peerreview]{IEEEtran}
\usepackage{hyperref} %
\usepackage%
{hyperref} %
\hypersetup{
    colorlinks,
    linkcolor={blue!80!black},%
    citecolor={green!30!black},
    urlcolor={blue!80!black}
}

\usepackage{balance}

\usepackage{amsmath,amssymb,amsfonts}
\usepackage{amsmath}
\usepackage{bbm}
\usepackage{nicefrac}
\usepackage{amsthm}

\usepackage{algorithmic}

\usepackage{textcomp}
\def\BibTeX{{\rm B\kern-.05em{\sc i\kern-.025em b}\kern-.08em
    T\kern-.1667em\lower.7ex\hbox{E}\kern-.125emX}}

\usepackage{physics}
\usepackage{siunitx}
\AtBeginDocument{\RenewCommandCopy\qty\SI}
\ExplSyntaxOn
\msg_redirect_name:nnn { siunitx } { physics-pkg } { none }
\ExplSyntaxOff

\usepackage{graphicx}
\usepackage{xcolor}

\usepackage{comment}

\renewcommand{\trace}{\mathrm{Tr}}
\AtBeginDocument{\RenewCommandCopy\qty\SI}

\theoremstyle{remark}	\newtheorem{theorem}{Theorem}
\theoremstyle{remark}	
\theoremstyle{remark}	\newtheorem{corollary}[theorem]{Corollary}
\theoremstyle{remark}	
\theoremstyle{remark} \newtheorem{definition}{Definition}
\theoremstyle{remark} \newtheorem{remark}{Remark}
\theoremstyle{remark} 

\usepackage[square,numbers,sort&compress]{natbib}
\usepackage[utf8]{inputenc}
\usepackage{mathrsfs}

\usepackage{accents}
\newlength{\dhatheight}

\usepackage{lscape}

\usepackage{tikz}
\usetikzlibrary{positioning}
\usetikzlibrary{decorations.markings}
\usetikzlibrary{arrows}
\usetikzlibrary{arrows.meta}
\usepackage[free-standing-units]{siunitx}
\usepackage[siunitx, RPvoltages]{circuitikz}

\makeatletter
\ctikzset{bipoles/my switch/height/.initial=.5}
\ctikzset{bipoles/my switch/width/.initial=.50}
\pgfcircdeclarebipole{}{}{myswitch}{%
\ctikzvalof{bipoles/my switch/height}}{\ctikzvalof{bipoles/my switch/width}}{
        \pgfsetlinewidth{\pgfkeysvalueof{/tikz/circuitikz/bipoles/thickness}\pgfstartlinewidth}
        \pgfpathmoveto{\pgfpoint{\pgf@circ@res@left}{0pt}}
        \pgfpathlineto{\pgfpoint{\pgf@circ@res@right}{.75\pgf@circ@res@up}}
        \pgfusepath{draw}
        \pgfsetlinewidth{\pgfstartlinewidth}        
        \pgftransformshift{\pgfpoint{\pgf@circ@res@left}{0pt}}
        \pgfnode{ocirc}{center}{}{}{\pgfusepath{draw}}
        \pgftransformshift{\pgfpoint{2\pgf@circ@res@right}{0pt}}
        \pgfnode{ocirc}{center}{}{}{\pgfusepath{draw}}
}
\def\pgf@circ@myswitch@path#1{\pgf@circ@bipole@path{myswitch}{#1}}
\compattikzset{my switch/.style = {\circuitikzbasekey, 
/tikz/to path=\pgf@circ@myswitch@path}}

\usepackage{xcolor}
\definecolor{mypurple}{rgb}{1,0,1}
	\definecolor{apricot}{rgb}{0.98, 0.81, 0.69}
		\definecolor{azure}{rgb}{0.0, 0.5, 1.0}
			\definecolor{darkmidnightblue}{rgb}{0.0, 0.2, 0.4}
				\definecolor{tearose}{rgb}{0.97, 0.51, 0.47}
					\definecolor{teagreen}{rgb}{0.82, 0.94, 0.75}
						\definecolor{indigo}{rgb}{0.29, 0.0, 0.51}
							\definecolor{tyrianpurple}{rgb}{0.4, 0.01, 0.24}

\tikzstyle{myedgestyle} = [-triangle 60]
\tikzstyle{block} = [draw, shape=rectangle, minimum height=3cm, minimum width=3cm, node distance=4cm, line width=0.5pt]
\tikzstyle{sum} = [draw, shape=circle, node distance=4cm, line width=0.5pt, minimum width=2.em]
\tikzstyle{mult} = [draw, shape=circle, node distance=3cm, line width=0.5pt, minimum width=2.em]
\tikzstyle{branch}=[fill,shape=circle,minimum size=4cm,inner sep=0pt]

\allowdisplaybreaks %to enable breaks within equations

\title{Coordination Capacity  for Classical-Quantum Correlations}
\author{% 
    \IEEEauthorblockN{Hosen Nator and Uzi Pereg} \\
    \IEEEauthorblockA{\normalsize Electrical and Computer Engineering and Helen Diller Quantum Center, Technion 
   }}
\begin{document}
\maketitle

\begin{abstract}
% THIS PAPER IS ELIGIBLE FOR THE STUDENT
% PAPER AWARD.

Network coordination  is considered in three basic settings, characterizing the generation of separable and classical-quantum correlations among multiple parties.
First, we consider the simulation of a classical-quantum state between two nodes with rate-limited common randomness (CR) and communication.  Furthermore, we study the preparation of a separable state between multiple nodes with rate-limited CR and no communication. At last, we consider a broadcast setting, where a sender and two receivers simulate a classical-quantum-quantum state using rate-limited CR and communication. 
We establish the optimal tradeoff between communication and CR rates in each setting.
%The objective of coordination addressed in this paper is to create a desired joint classical-quantum state between the different parts of a given system using only classical communication and common randomness. 
% Cuff (2008) found the coordination capacity region for a point-to-point classical system  and \mbox{Cuff et al.} (2010) for the three-node no-communication network. 
%We generalize the results to classical-quantum systems.
\end{abstract}

\vspace{0.1cm}

\section{Introduction}
Coordination is essential in communication systems, as it ensures that different components can work together in 
harmony to achieve a common goal. For example,  in sensor networks,  the sensors do not just share information in the conventional sense, but also collaborate in the transmission of  data \cite{sensor_coordination_2011}.
% cooperation_quantum_sensor_network_2019}. 
Furthermore, coordination plays a major role in cooperative computing between distributed systems \cite{Coordination_in_distributed_computing_2002},  %distributed task management \cite{Distrebuted_task_management_2018} and 
%task-oriented %communication  in 
%cyber-physical communication 
%\cite{Task_oriented_cyber_2023}, 
rate-distortion theory for secrecy systems \cite{Rate_distortion_2014}, %distributed channel synthesis \cite{Distributed_Channel_Synthesis_2013}, 
simulation of distributed quantum measurements \cite{Faithful_simulation_Heidari_2019}, and quantum nonlocal games \cite{kaur2020nonlocality}.
 It is envisioned that quantum information technology will
enhance %and elevate 
future communication systems
from different perspectives, such as
efficiency \cite{fettweis20226g,nemirovsky2023increasing}, security \cite{QKD_secrecy_2020,lederman2024secure,PeregDeppeBoche:21p}, and computing \cite{6G_quantum_2022}.
These advances motivate the study of
  coordination in quantum networks \cite{BBDFFJS:21b}. 

Cuff et al. \cite{%cuff2008communication,
cuff2010coordination} studied classical coordination in various communication networks with different topologies, and considered two coordination types, empirical coordination and strong coordination. Empirical coordination requires that the average frequency of joint actions in the network approaches a desired distribution with high certainty. On the other hand, strong coordination sets a requirement on  the joint distribution of all actions.  %The latter is a stricter condition on the joint distribution of the nodes, and not just the average.
Efficient coordination codes are constructed in \cite{strong_emprical_2018}.
Source coding \cite{Compressing_mixed_state_sources_2002, Classical_broadcast_cooperation_2016,Quantum_Classical_Source_Coding2023,Rate_limited_source_coding_2023}, %\textcolor{blue}{should I add the rated distortion papers here? I am refering to \cite{Quantum_Classical_Source_Coding2023,Rate_limited_source_coding_2023} }, 
 state coordination \cite{Multiple_access_channel_simulation_2011},
channel simulation \cite{%cuff2008communication,
channelsimulation2024,
berta2013entanglement,
BennettDevetakHarrowShorWinter:14p,pirandola2018theory,wilde2018entanglement,%li2021reliable_channel_simulation,
cheng2023quantum}, 
and distributed 
source simulation \cite{Wyner_distributed_source_simulation_2018,state_generation_using_correlated_resource_2023}
can be viewed as instances of network coordination. %between two nodes.
% Cuff \cite{cuff2008communication} solved the classical channel simulation problem  under the strong coordination requirement.
  In particular,
  Bennet et al. \cite{BennettDevetakHarrowShorWinter:14p} considered   simulation of a quantum channel, under the assumption that pre-shared entanglement is available to the sender and the receiver,
and derived the quantum reverse Shannon theorem \cite{abeyesinghe2009mother,Berta_reverse_Shannon_2020}. 
The optimal simulation rate turns out to be identical to the entanglement-assisted 
 quantum capacity %for the transmission of classical information over a quantum channel 
 \cite{BennettDevetakHarrowShorWinter:14p}.
Recently, the authors \cite{NatorPereg:24c1} considered
entanglement coordination using quantum communication links.

   The  general problem of quantum coordination can be formulated as follows. Consider a quantum network that consists of  $m$ nodes, where Node $i$ performs an encoding operation $\mathcal{E}_i$ on a quantum system  $A_i$, which is required to be in a certain desired correlation  with the rest of the network. In other words, the goal is to simulate a particular joint state, $\omega_{A_1 A_2 \cdots A_m}$.
   In general, some of the nodes are not free to choose their encoding, but rather their state is dictated by Nature, according to a certain physical process.
 Node $i$ can also send a sequence of bits or qubits to Node $j$ via  a communication link of a limited rate, $R_{i,j}$.  %Given a  joint density operator $\omega_{A_1,A_2,\dots,A_m}$, 
The ultimate performance is defined by %The object of interest is %to determine 
the set of rates $\{R_{i,j}\}%_{1\leq i,j\leq m}
 $ that are necessary and  sufficient in order to simulate the quantum correlation.  
 % The quantum coordination capacity region is defined as the collection of all such rate matrices $\{R_{i,j}\}_{1\leq i,j\leq m}$
 % in $\mathbb{R}^{m\times m}$. 

Cuff et al. \cite{%cuff2008communication,
cuff2010coordination} introduced the  classical version of this problem,
where the encoders, decoders, and rates are all classical, and the goal is to simulate a prescribed probability distribution. %And qubits are replaced with bits.
In the basic two-node setting, the simulation of a joint distribution $p_{XY}$ requires a rate
 $R_{1,2}\geq C(X;Y)$, where $C(X;Y)$  is Wyner's common information \cite{wyner1975common}.
 The quantum analog  was recently established by George et al. \cite{state_generation_using_correlated_resource_2023}, in the context of distributed source simulation.
 Under the assumption that Alice and Bob share
unlimited % a sufficient amount of 
common randomness (CR) a priori,  simulation can be performed at a lower rate,  $R_{1,2} \geq I(X;Y)$, where $I(X;Y)$ is the mutual information. %with respect to the joint distribution $p^*(x,y)$.
 The capacity region describes the optimal tradeoff between the communication and CR rates  \cite{cuff2008communication}. 

In this paper, we consider three coordination settings. Our models are motivated by quantum-enhanced Internet of Things (IoT) networks in which the communication links are classical \cite{notzel2020entanglement,9232550,burenkov2021practical,granelli2022novel}, as opposed to our work in \cite{NatorPereg:24c1}. 
% % of coordination between nodes in . 
First, we consider the simulation of a classical-quantum state $\omega_{XB}$ between two nodes with rate-limited CR.
% First, we consider
% coordination in a two-node network in order to simulate
%  a classical-quantum (c-q) state $\omega_{XB}$. 
%
We characterize the optimal tradeoff
between the required rate of description and the amount of CR used.
 %
 % In this setting, the first node, which corresponds to Alice, receives a classical sequence $X^n$ drawn i.i.d. from a classical source with a PMF $p_X$. The second node corresponds to Bob, who has access to a quantum system $B$. Both nodes can communicate classically at a rate of $R_1$. Moreover,  both users have access to shared random bits at a rate $R_0$. We would like to find the constraints on the rate pairs $\left(R_0, R_1\right)$ so that the joint state between Alice and Bob is $\omega_{XB}$. We found that the constraints are $R_1>I(X;U)$ and $R_0+R_1>I(X,B;U)$, where $U$ is a helping random variable which satisfies that given $U$, system $B$ is independent of $X$. 
%
Our second model  is a quantum  %generalization of coordination in a 
no-communication network. %problem  \cite{cuff2010coordination}. 
The network comprises three nodes, where no-communication is allowed between the nodes, yet CR is available  at a classical rate $R_0$. 
Thereby, only separable states can be simulated.
%We determine the optimal CR rate for simulation of joint state, which mu. % $R_0$. 
We show that a joint  state $\omega_{ABC}$ can be simulated at a CR rate of $R_0\geq I(ABC;U)_\sigma$, with respect to an extension $\sigma_{UABC}$ of $\omega_{ABC}$, where $U$ is an auxiliary classical random variable that satisfies a Markov property. % which satisfies that given $U$, systems $A,B\text{ and }C$ are independent. 
At last, we consider a broadcast setting, where a sender and two receivers simulate a classical-quantum-quantum (c-q-q) state using rate-limited CR and communication. 
We establish the optimal tradeoff between communication and CR rate.
%The third setting is a broadcast network, where we simulate a classical-quatnum-quantum (c-q-q) state $\omega_{XB_1B_2}$ between one sender and two receivers. 
In the analysis, we use random coding and apply quantum resolvability results \cite{winter2001compression,hayashi2016quantum,Bloch_resolvability_2019}. %implied by the resolvability work of Han and Verdú in \cite{Han_and_Verdu_1993}. However, the concept was first introduced by Wyner in Theorem 6.3 in \cite{wyner1975common} by Wyner. We also use quantum resolvability results implied by lemma 5.2 in \cite{Bloch_resolvability_2019} by Bloch et al, which is based on the quantum channel resolvability in \cite{hayashi2016quantum}. 

\section{Problem Definitions}
We consider three coordination settings as described below. %Each one of the problems is addressed in a separate subsection.
%\subsection*{Notation}
%\label{Subsec:Notation}
We use standard notation in quantum information theory,  as in
\cite%[Chap. 11]
{wilde2017quantum},
$X,Y,Z,\ldots$ are discrete random variables on finite alphabets   $\mathcal{X},\mathcal{Y},\mathcal{Z},...$, respectively, 
 % The distribution of  $X$ is specified by a probability mass function (pmf) 
	% $p_X(x)$ on $\mathcal{X}$. %The set of all pmfs over $\Xset$ is denoted by $\pSpace(\Xset)$.
	% %	
% We use 
$x^n=(x_i)_{i\in [n]}$ denotes %for  
a sequence in %of letters from  
 $\mathcal{X}^n$. 
A quantum state %of a quantum system $A$ 
is described by 
a density operator, $\rho_A$, on the Hilbert space $\mathcal{H}_A$.
Denote the set of all such %density 
operators  by $\mathfrak{S}(\mathcal{H}_A)$.
A c-q channel is a map $\mathcal{N}_{X\to B}:\mathcal{X}\to \mathfrak{S}(\mathcal{H}_B)$.
%
%A measurement  is specified by a collection of operators $\{D_j \}$ that forms a POVM. % positive operator-valued measure (POVM). %, i.e.,
% $D_j\geq 0$  and  
%$\sum_j D_j=\identity$, where $\identity$ is the identity operator.
 % The qubit Pauli basis is denoted by $\{\identity,\mathsf{X},\mathsf{Y},\mathsf{Z}\}$.
%Given a bipartite state $\rho_{AB}$, %on $\mathcal{H}_A\otimes \mathcal{H}_B$,  
%define 
The quantum mutual information is defined as
$%\begin{align}
I(A;B)_\rho=H(\rho_A)+H(\rho_B)-H(\rho_{AB}) 
$, %\end{align} 
where $H(\rho) \equiv -\trace[ \rho\log(\rho) ]$, % is the von Neumann entropy,
the conditional quantum entropy as 
$H(A|B)_{\rho}=H(\rho_{AB})-H(\rho_B)$,  and
$I(A;B|C)_{\rho}%=H(A|C)_\rho+H(B|C)_\rho-H(A,B|C)_\rho$
$, %is defined 
accordingly. %respectively.

\subsection{Two-Node Network} 
\begin{figure}[bt]
\center
\includegraphics[scale=0.75,trim={5.3cm 0 5.5cm 0}]
{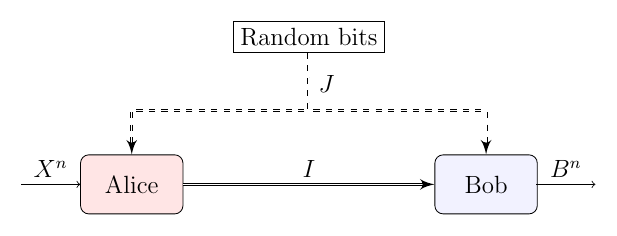} %  [trim={left bottom right top},clip]
\caption{Two-node network.
}
\label{Figure 1: Coordination capacity - two nodes}
\end{figure}

Consider the two-node network in Figure~\ref{Figure 1: Coordination capacity - two nodes}.
Alice and Bob wish to simulate a c-q state $\omega_{XB}^{\otimes n}$, using the following scheme.
Node 1 (Alice) %belongs to Alice
%and 
receives a classical source sequence $x^n$, drawn by Nature according to a given PMF $p_{X}$. The source sequence is encoded into an index $i$ at a rate $R_{1}$.
%by $I\in\mathcal{I}\triangleq\left[2^{nR_{1}}\right]$. 
%
Node 2 (Bob) is quantum. % belongs to Bob and corresponds to a quantum system $B$. 
Both
nodes have access to a CR element $j$ at a given rate $ R_0$, i.e., $j$ is
uniformly distributed over $\left[2^{n R_0}\right]$,
and it is independent of $X^{n}
$. 

\begin{figure}[bt]
\center
\includegraphics[scale=0.8,trim={3.75cm 0 4cm 0}]
{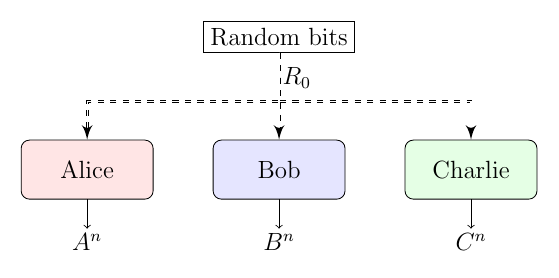} %  [trim={left bottom right top},clip]
\caption{%Coordination setting for the 
No-communication network}
\label{Figure 2 - No-communication network}
\end{figure}

Formally, 
%\begin{definition}
a $\left(2^{nR_{0}},2^{n R_1},n\right)$
coordination code for the 
simulation of a c-q state 
$\omega_{XB}
$
%two-node network %in Figure~\ref{Figure 1: Coordination capacity - two nodes} 
consists
of a classical encoding channel,
$%\begin{align}
F:\mathcal{X}^{n}\times [2^{n R_0}]  \to [2^{nR_1}] %\,,
$, %\end{align}
and a  c-q decoding channel %,  
$%\begin{align}
\mathcal{D} _{IJ\to B^n} %\,.
$. %\end{align}
%\end{definition}
%
The %coordination 
protocol works as follows.
A classical sequence
$x^n\sim p_X^n$ is generated by Nature. 
% begins by generating a random classical codebook $\mathcal{C}^n=\left\{{u^{n}\left(i,j\right)}\right\}_{(i,j)\in\mathcal{I}\times\mathcal{J}}$ generated according to a PMF $p_{U}^n=\prod_{i=1}^{n}p_U(u_i)$. The alphabet $\mathcal{U}$ is the support of $p_U$, the required PMF $p_U$ \textcolor{blue}{will be determined later in this subsection}. The codewords are a function of both the
% encoded source $I$ and the common randomness $J$. 
%Once the codebook is generated, it is shared with both Alice and Bob. 
% Given the input $x^n$ and the CR element $j$, Alice encodes a message $i=F (x^n,j)$, which she sends to Bob via a noiseless classical link at a rate $R_1$. Upon receiving $i$, Bob  uses his  c-q map in order to prepare a quantum state
% $\mathcal{D}_{n}\left(i,j\right)=\bar{\rho}_{B^n}^{U(i,j)}$. Denote the joint state of Alice and Bob after the end of the coordination protocol by $\widehat{\rho}_{X^nB^n}$. We want the state $\widehat{\rho}_{X^nB^n}$ to be arbitrarily close to the desired state $\omega_{XB}^{\otimes n}$.
%
%In addition, the common randomness element $j$ is chosen uniformly at random, from $[2^{n R_0}]$.
Given the sequence $x^n$ and the CR element $j$, 
Alice selects a random index,
\begin{align}
i\sim
F(\cdot|x^{n},j) %\,.
\end{align}
and sends it through a noiseless link.
As Bob receives the message $i$ and the CR element $j$, he prepares the state
\begin{align}
\rho_{B^n}^{(i,j)}=\mathcal{D}_{IJ\to B^n} (i,j) \,. 
\end{align}
Hence, the resulting joint state is
\begin{align}
% &
\widehat{\rho}_{X^n B^n}=\frac{1}{2^{n R_0}}\sum_{j\in [2^{n R_0}]}
\sum_{x^n\in\mathcal{X}^n}\bigg( p_X^n(x^n) \ketbra{x^n}_{X^n}
% \nonumber\\
% &
\otimes \sum_{i\in [2^{nR_1}]} F (i|x^n,j) \rho_{B^n}^{(i,j)} \bigg) \,.
\end{align}

\begin{definition}
\label{Definition_2}
A coordination rate pair $\left(R_0,R_1\right)$
is %said to be 
achievable for the simulation of $\omega_{XB}$, if for every $\varepsilon>0$ and sufficiently large $n$,
there exists a  $\left(2^{nR_0},2^{nR_1},n\right)$
%coordination 
code that achieves
\begin{align}
\label{Equation:Error_2}
\norm{\widehat{\rho}_{X^nB^n}-\omega_{XB}^{\otimes n}}_1 \leq \varepsilon  \,.
\end{align}
%\end{definition}
%
%\begin{definition}
 The coordination
capacity region of the two-node network, $\mathcal{R}_{\text{2-node}}(\omega)$, with respect to the c-q state $\omega_{XB}$, is the closure of the set of all achievable rate pairs. % $\left(R_{0},R_{1}\right)$.

The coordination capacity,
$\mathsf{C}_{\text{2-node}}^{(0)}(\omega)$, without CR, is the supremum of rates $R_1$ such that $\left(0,R_{1}\right)\in \mathcal{R}_{\text{2-node}}(\omega)$.
The CR-assisted coordination capacity,
$\mathsf{C}_{\text{2-node}}^{(\infty)}(\omega)$,  i.e., with unlimited CR, is the supremum of rates $R_1$ such that $\left(R_{0},R_{1}\right)\in \mathcal{R}_{\text{2-node}}(\omega)$ for some $R_1\geq 0$.

 \label{Definition:Two_Node_Coordination_Region}  
\end{definition}

\subsection{No-Communication Network}
Consider a network that consists of three users: Alice, Bob and Charlie, holding %each has a 
quantum systems  $A$, $B$, and $C$, respectively. The users cannot communicate, but they share a CR element $j$  at a  rate $R_0$, as illustrated in Figure ~\ref{Figure 2 - No-communication network}. Given $j$, each  user prepares a quantum state  separately. 
% We seek to identify the needed rate $R_0$ needed in order to create a joint quantum state $\omega_{ABC}^{\otimes n}$.
% The coordination protocol begins by generating a random classical codebook $\mathcal{F}^n=\left\{V^n\right\}_{l\in\mathcal{L}}$ according to a PMF $p^n_V(v^n)=\prod_{k=1}^np_V(v_k)$, where $p_V$ has a support $\mathcal{V}$. After the codebook is generated, it is revealed to Alice, Bob and Charlie. 
% Given $L$ each user finds the codeword $V^n(L)$ and prepares a quantum state, Alice prepares a state $\Bar{\sigma}_{A}^{V^n(L)}$, Bob prepares a state $\Bar{\sigma}_{B}^{V^n(L)}$ and Charlie prepares a state $\Bar{\sigma}_{C}^{V^n(L)}$. Denote the final state at the end of the protocol by $\widehat{\Sigma}_{A^nB^nC^n}$. We want the state $\widehat{\Sigma}_{A^nB^nC^n}$ to be arbitrarily close to $\Omega_{ABC}^{\otimes n}$.

%\begin{definition}
A $\left(2^{nR_{0}},n\right)$
coordination code for the no-communication network 
%in Figure~\ref{Figure 2 - No-communication network} 
consists of a CR set  $[2^{nR_0}]$, and three  c-q encoding channels,
%\begin{align}
    $\mathcal{T}^{(1)} _{J\to A^n}$, % \,,\;
    $\mathcal{T}^{(2)} _{J\to B^n}$, and % \,,\;
    $\mathcal{T}^{(3)} _{J\to C^n}$.  %\,.
%\end{align}
%The decoding channels are assoicated Alice, Bob, and Charlie respectively.
%\end{definition}
As Alice, Bob, and Charlie receive a realization $j$ of the CR element, each uses their encoding map to prepare their respective state. 
prepares a quantum state, 
%\begin{align}
$
    \rho_{A^n}^j=  \mathcal{T}^{(1)} _{J\to A^n}(j)%\\
$,  
$
    \rho_{B^n}^j=  \mathcal{T}^{(2)} _{J\to A^n}(j)%\\
$, and
$
    \rho_{C^n}^j=  \mathcal{T}^{(3)} _{J\to C^n}(j)
$, respectively. 
Hence,
%\end{align}
%This results in the joint state,
\begin{align*}
&\widehat{\rho}_{A^n B^n C^n}=\frac{1}{2^{n R_0}}\sum_{j\in [2^{n R_0}]} \mathcal{T}^{(1)}% _{J\to A^n}
(j) 
\otimes \mathcal{T}^{(2)} %_{J\to B^n}
(j)  \otimes \mathcal{T}^{(3)} %_{J\to C^n}
(j) \;.
\end{align*}

\begin{definition}
A CR rate $R_{0}$
is %said to be 
achievable for the simulation of $\omega_{ABC}$, if for every $\varepsilon>0$ and sufficiently large $n$,
there exists a  $\left(2^{nR_{0}},n\right)$
coordination code that achieves
\begin{align}\norm{\widehat{\rho}_{A^n B^n C^n}-\omega_{A B C}^{\otimes n}}_1 \leq \varepsilon  \,.
\end{align}
%\end{definition}
%
%\begin{definition}
 The coordination
capacity $\mathsf{C}_{\text{NC}}(\omega)$, for the no-communication network,  is the infimum of achievable rates $R_0$.
If there are no achievable rates, we 
set $\mathsf{C}_{\text{NC}}(\omega)=+\infty$.
\end{definition}

\subsection{Broadcast Network}
\begin{figure}[bt]
\center
\includegraphics[scale=0.85,trim={5.8cm 0 5.5cm 0}]
{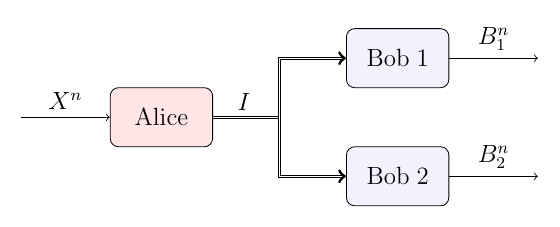} %  [trim={left bottom right top},clip]
\caption{Broadcast Network. %Coordination setting for the broadcast network. 
The CR element is omitted for simplicity. %available to all nodes of the network, however we omitted it for convenience.
}
\label{Figure 3: Broadcast_network}
\end{figure}

Consider the broadcast network in Figure~\ref{Figure 3: Broadcast_network}. A sender, Alice,  and two receivers, Bob~1 and Bob~2, wish to simulate a c-q-q state $\omega_{XB_{1}B_{2}}$, using the following scheme. Alice receives a classical source sequence $x^{n}\in\mathcal{X}^{n}$ drawn by Nature, i.i.d. according to a given PMF $p_{X}$. Alice encodes the source sequence into an index $i$ at a rate $R_{1}$. The other two nodes, of Bob~1 and Bob~2, are quantum. The three nodes have access to a CR element $j$ at a rate $R_{0}$. %, i.e., $j$ is uniformly distributed over $[2^{nR_{0}}]$ and  independent of $x^{n}$.
Similarly, a $\left(2^{nR_{0}},2^{nR_{1}},n\right)$ coordination code %for the simulation of a c-q-q state $ \omega_{XB_{1}B_{2}}$ 
consists of a classical encoding channel, 
$%\begin{align}
F:\mathcal{X}^{n}\times[2^{nR_{0}}]\to[2^{nR_{1}}]\,,
$ %\end{align}
and two c-q decoding channels,
%\begin{align}
    $\mathcal{D}^{(\ell)}_{IJ\to B_{\ell}^{n}}$,  %\,\text{ 
    for $\ell\in\{1,2\}$. %}\, %\mathcal{D}^{(2)}_{IJ\to B_{2}^{n}}\,
    %.
%\end{align}
%The coordination protocol operates in the following way. 
Given %the classical source sequence 
$x^{n}$ and the CR element $j$, Alice generates %an index %$i\in[2^{nR_{1}}] $,
$%\begin{align}
    i\sim F(\cdot|x^{n},j)
$, %\end{align}
and sends it to both Bob~1 and Bob~2, who then %. Then, each 
apply their decoding map.
% , preparing
% %they prepare the following states respectively
% % \begin{align}
% %     \rho_{B_{1}^{n} B_2^n}^{(i,j)}=\mathcal{D}_{IJ\to B_{1}^{n}}(i,j)\otimes \mathcal{D}_{IJ\to B_{2}^{n}}(i,j)\,.
% % \end{align}
% % Therefore, the resulting joint state is 
% \begin{multline}
%     \widehat{\rho}_{X^{n}B_{1}^{n}B_{2}^{n}}	=\frac{1}{2^{nR_{0}}}\sum_{j\in[2^{nR_{0}}]}\sum_{x^{n}\in\mathcal{X}^{n}}\bigg(p_{X}^{n}(x^{n})\ketbra{x^{n}}_{X^{n}} 
%     \\ 
%     \otimes\sum_{i\in[2^{nR_{1}}]}F(i|x^{n},j)
%     \mathcal{D}^{(1)}_{IJ\to B_{1}^{n}}(i,j)\otimes \mathcal{D}^{(2)}_{IJ\to B_{2}^{n}}(i,j)
%     %\rho_{B_{1}^{n} B_2^n}^{(i,j)}%\otimes\rho_{B_{2}^{n}}^{(i,j)}
%     \bigg)\,.
%\end{multline}

% \begin{definition} 
% A coordination rate pair $\left(R_{0},R_{1}\right)$ is said to be achievable for the simulation of $\omega_{XB_1 B_2}$, if for every $\varepsilon>0$ and sufficiently large n, there exists a $\left(2^{nR_{0}},2^{nR_{1}},n\right)$ coordination code that achieves 
% \begin{align}
% \norm{\widehat{\rho}_{X^{n}B_{1}^{n}B_{2}^{n}}-\omega_{XB_{1}B_{2}}^{\otimes n}}_{1}\leq\varepsilon
% \end{align} 
% % \end{definition}
% %
% % \begin{definition}
The coordination capacity region of the broadcast network, $\mathcal{R}_{\text{BC}}(\omega)$, with respect to the c-q state $\omega_{XB_{1}B_{2}}$, is defined in a similar manner as in Definition~\ref{Definition_2}. % the closure of the set of all achievable rate pairs $\left(R_{0},R_{1}\right)$.
%\end{definition}

\section{Results}
\subsection{Two-Nodes Network}
Consider a given c-q state $\omega_{XB}$ that we wish to simulate. We now state our main result.
Define the following set of c-q states.
Let $\mathscr{S}_{\text{2-node}}(\omega)$ be the set of all c-q states
\begin{subequations}
\begin{align}
  \sigma_{XUB}&=
 \sum\limits_{
 \substack{
 (x,u)\in\\
 \mathcal{X}\times\mathcal{U}}
 }
 p_{X,U}(x,u)\ketbra{x}_{X} 
 %\\
 \; \otimes
 \ketbra{u}_{U}\otimes \theta_{B}^{u} 
\end{align}
such that %\intertext{such that}
\begin{align}
 \sigma_{XB}&=\omega_{XB}
\end{align}
\end{subequations}
% and 
% \begin{align}
%  \rho_{B}^{(x,u)}=\bar{\rho}_{B}^{u}  \,,\;\text{for $x\in\mathcal{X}$} \,.
% \end{align}
for
$%\begin{align}
|\mathcal{U}|\leq |\mathcal{X}|^2[\mathrm{dim}(\mathcal{H}_B)]^2+1
$. %\end{align}
Notice that given a classical value $U=u$, there is no correlation between $X$ and $B$.

\begin{theorem}
\label{Theorem_2}
The coordination capacity region for the two-node system described in Figure~\ref{Figure 1: Coordination capacity - two nodes} is the set
\begin{align}
\mathcal{R}_{\text{2-node}}(\omega)=
\bigcup_{%\sigma_{XUB}\in 
\mathscr{S}_{\text{2-node}}(\omega)}
\left\{ 
\begin{array}{l}
\left(R_{0}, R_{1}\right)\in\mathbb{R}^{2}:
\\
%\exists\rho_{XUB}\in \mathscr{S}\;s.t.\\
R_{0}        \geq I(X;U)_\sigma \,,\\
R_0+R_1  \geq I(XB;U)_\sigma
\end{array}
\right\} \,.
\end{align}
% where 
% \begin{align}
%   D\triangleq 
%   \left\{
%   \begin{array}{l}
%   \rho_{XUB}=
%  \sum\limits_{(x,u)\in\mathcal{X}\times\mathcal{U}}p_{X,U}^*(x,u)\ketbra{x}_{X} 
%  \\
%  \; \otimes
%  \ketbra{u}_{U}\otimes\rho_{B}^{(x,u)} 
% \\
%  \rho_{XB}=\omega_{XB}
% \\
%  \rho_{B}^{(x,u)}=\bar{\rho}_{B}^{u}  \;\forall x\in\mathcal{X}
%  \end{array}
%  \right\} 
% \end{align}

\end{theorem}
% Please notice that in group $D$, given $U$, system $B$ is independent of $X$ therefore
% $\rho_{B}^{x,u}=\bar{\rho}_{B}^{u}$. 
The proof for Theorem~\ref{Theorem_2} is given in Subsection~\ref{Subsection:Proof_2}.
The following corollaries immediately follow.

\begin{corollary}[Quantum  Common Information
\cite{state_generation_using_correlated_resource_2023}] The coordination capacity without CR is
\begin{align}
\mathsf{R}_{\text{2-node}}^{(0)}(\omega)=
\min_{\sigma_{XUB}\in \mathscr{S}_{\text{2-node}}(\omega)}
 I(XB;U)_\sigma  \,.
\end{align}

\end{corollary}

\begin{corollary}
The CR-assisted coordination capacity, i.e., with unlimited common randomness, is given by
\begin{align}
\mathsf{R}_{\text{2-node}}^{(\infty)}(\omega)\triangleq
\min_{\sigma_{XUB}\in \mathscr{S}_{\text{2-node}}(\omega)}
%\exists\rho_{XUB}\in \mathscr{S}\;s.t.\\
 I(X;U)_\sigma %\,,\\
%R_0  \geq I(U;B|X)_\sigma
\end{align}
\end{corollary}
We note that in order to achieve the CR-assisted capacity, a CR rate of $R_0=I(U;B|X)_{\sigma}$
is sufficient.
If $B\equiv Y$ is classical, then
%In the classical-classical case of $\omega_{XY}$, where $Y$ is classical,
 we may substitute $U=Y$, which yields  the capacity $\mathsf{R}_{\text{2-node}}^{(\infty)}(\omega)=I(X;Y)$, and it can be achieved with CR at rate $R_0=H(Y|X)$  \cite{cuff2008communication}.

\subsection{No-Communication Network}

Consider a given quantum state $\omega_{ABC}$ that we wish to simulate. We now state our main result.
Define the following set of c-q states.
Let $\mathscr{S}_{NC}(\omega)$ be the set of all c-q states
\begin{subequations}
 \label{Equation:S_NC}
\begin{align}
  \sigma_{UABC}&=
 \sum\limits_{u\in\mathcal{U}}p_{U}(u)
 %\\
 \; 
 \ketbra{u}_{U}\otimes \theta_{A}^{u}\otimes\theta_{B}^{u}\otimes\theta_{C}^{u} 
\end{align}
such that %\intertext{such that}
\begin{align}
 \sigma_{ABC}&=\omega_{ABC}
\end{align}
\end{subequations}
% for
% \begin{align}
% |\mathcal{U}|\leq [\mathrm{dim}(\mathcal{H}_A) \mathrm{dim}(\mathcal{H}_B) \mathrm{dim}(\mathcal{H}_C)]^2
% \end{align}
%
Given $U=u$, there is no correlation between $A,B$ and $C$.

\begin{theorem}
\label{Theorem:NC}
The coordination capacity  for the no-communication network described in Figure~\ref{Figure 2 - No-communication network} is %the set
\begin{align}
\mathsf{C}_{\text{NC}}(\omega)=
\inf_{\sigma_{UABC}\in\mathscr{S}_{NC}(\omega)}
%\exists\rho_{XUB}\in \mathscr{S}\;s.t.\\
 I(U;ABC)_\sigma
\end{align}
with %where we use 
the convention that an infimum over an empty set  is $+\infty$.
\end{theorem}

\begin{remark}
\label{Remark:Entanglement_2}
Since the CR is classical, it cannot be used in order to create entanglement. 
Therefore, as Alice, Bob, and Charlie do not cooperate with one another, 
 it is impossible to simulate entanglement.
 That is, we can only simulate separable states. 
\end{remark}

% As pointed out in Remark~\ref{Remark:Entanglement_2}, the joint state cannot be entangled, as there is no cooperation or quantum correlation to begin with.
% In other words, if $\omega_{ABC}$ is entangled, then $\mathsf{C}_{\text{NC}}(\omega)=+\infty$.

\subsection{Broadcast Network}
\label{Subsection:BC}
Consider a given c-q-q state $\omega_{XB_{1}B_{2}}$ that we wish to simulate. 
Define the following set of c-q-q states. Let $\mathscr{S}_{\text{2-BC}}(\omega)$ be the set of all c-q states
%\begin{subequations}
    \begin{align*}
    \sigma_{XUB_{1}B_{2}}	&=\sum\limits _{\substack{(x,u)\in \\ \mathcal{X}\times\mathcal{U}}}p_{XU}(x,u)\ketbra{x}_{X}
  %  \nonumber\\ &
  \otimes\ketbra{u}_{U}\otimes\theta_{B_{1}}^{u}\otimes\eta_{B_{2}}^{u}
\end{align*}
such that %\intertext{such that}
\begin{align*}
 \sigma_{XB_1 B_2}&=\omega_{XB_1 B_2} \,.
\end{align*}
%\end{subequations}
% and 
% \begin{align}
%  \rho_{B}^{(x,u)}=\bar{\rho}_{B}^{u}  \,,\;\text{for $x\in\mathcal{X}$} \,.
% \end{align}
% for
% \begin{align}
% |\mathcal{U}|\leq |\mathcal{X}|^2[\mathrm{dim}(\mathcal{H}_B)]^2+1
% \end{align}
%
 Note that given %a classical value 
% $U=u$,  
$X$, $B_1$, and $B_2$ are uncorrelated given %a classical value 
$U=u$. 
\begin{theorem}
\label{Theorem: Broadcast}
The coordination capacity region of the broadcast channel in Figure~\ref{Figure 3: Broadcast_network} network is the set
\begin{align}%{multline}
\mathcal{R}_{\text{BC}}(\omega)=%\\
\bigcup_{%\sigma_{XUB_{1}B_{2}}\in
\mathscr{S}_{BC}(\omega)}
\left\{ 
\begin{array}{l}
(R_0,R_1)\in\mathbb{R}^2:
\\
R_0  \geq I(X;U)_\sigma
\\
R_0+R_1  \geq
%\\{\;\;\;\;\;\;\;\;\;}
I(XB_1B_2;U)_\sigma
\end{array}
\right\} \,.
\end{align}%{multline}
\end{theorem}

\begin{remark}
\label{Remark:Entanglement_BC}
Since Alice's encoding is classical, she cannot distribute entanglement. 
Therefore, as Bob~1 and Bob~2 do not cooperate with one another, 
 it is impossible to simulate entanglement between Bob~1 and Bob~2.
 That is, we can only simulate states such that $\omega_{B_1 B_2}$ is separable, as in the no-communication model (see Remark~\ref{Remark:Entanglement_2}). 
\end{remark}
%In a similar manner as in Remark~\ref{Remark:Entanglement_2},
% Since Alice's encoding is classical, she cannot distribute entanglement. 
% the joint state cannot be entangled, as there is no cooperation or quantum correlation to begin with.
% In other words, if $\omega_{ABC}$ is entangled, then $\mathsf{C}_{\text{NC}}(\omega)=+\infty$.
% % \begin{corollary}
    
% \end{corollary}

\section{Two Node Analysis}
%\section{Proofs}
%\subsection{Two-nodes with communication}
\label{Subsection:Proof_2}
%
%In this section, we prove Theorem~\ref{Theorem_2}. 
Consider the two node network in Figure~\ref{Figure 1: Coordination capacity - two nodes}.
Our proof for Theorem~\ref{Theorem_2} is based on %random coding and 
quantum resolvability %results 
\cite{winter2001compression,hayashi2016quantum,Bloch_resolvability_2019}. %combines the tools of classical resolvability in a similar way to \cite{cuff2008communication}, and quantum resolvability as applied in lemma 3 in \cite{Bloch_resolvability_2019} based on 
%\cite{hayashi2016quantum}.

% \begin{theorem}
% Let $p_{U,V}\left(u,v\right)$ be a discrete distribution. For any
% $n\in\mathbb{N}$, define the distributions $p_{U}^{n}\left(u^{n}\right)=\prod_{k=1}^{n}p_{U}\left(u_{k}\right)$.
% Let\textbf{ }$\mathcal{C} =\left\{ u^{n}(m)\right\} _{m=1}^{2^{nR}}$
% be a codebook of sequences drawn i.i.d. according to $p_{U}^{n}\left(u^{n}\right)$.
% Once the codebook is generated, define the following distribution
% \begin{align}
%     Q\left(v^{n}\right)=\frac{1}{2^{nR}}\sum_{m=1}^{2^{nR}}\prod_{k=1}^{n}p_{V|U}^{n}\left(v_{k}|U_{k}(m)\right)
% \end{align}
% Cuff in \cite{cuff2008communication} showed that using classical resolvability that the following holds
% If $R>I\left(V;U\right)$ then,
% \begin{align}
%     \lim_{n\to\infty}\mathbb{{E}}_{\mathcal{C} }\left[\left\Vert Q\left(v^{n}\right)-\prod_{i=1}^{n}p_{V}(v_{k})\right\Vert _{1}\right]=0
% \end{align}
% \end{theorem}
\begin{theorem}[see \cite{winter2001compression,hayashi2016quantum,Bloch_resolvability_2019}]
\label{Theorem:Quantum_resolvability}
Consider an ensemble, $\{ p_X, \rho_A^x \}_{x\in\mathcal{X}}$, and %.
%Furthermore,
%consider 
a random codebook that consists of $2^{nR}$ independent sequence,
%Any random encoder $U:\left[2^{nR}\right]\rightarrow\mathcal{U}^{n}$
%with a codebook $\mathcal{C} $ which consists of codewords 
$ X^{n}(m)$, $m\in [2^{nR}]$,  each is i.i.d. %according to the distribution 
$\sim p_{X}%\left(u\right)
$.
If $R>I(X;A)_{\rho}$, then for every $\delta>0$ and sufficiently large $n$,
%there exists $n_{1}(\varepsilon)$ such that for every
%$n>n_{1}(\varepsilon)$ the following holds:
%
\begin{align}
\mathbb{E}%_{\mathcal{C} }
\left[
\norm{
\rho_{A}^{\otimes n}-
\frac{1}{2^{nR}}
\sum_{m=1}^{2^{nR}} \rho_{A^n}^{X^{n}(m)}
}_{1}\right]\leq\delta \,,
\label{Equation:Quantum_resolvability}
\end{align}
where $\rho_{A^n}^{x^n}\equiv \bigotimes_{k=1}^n \rho_A^{x_k}$, 
and the expectation is %calculated 
over all realizations of the random codebook. % consisting
%of $2^{nR}$ sequences of length $n$ drawn i.i.d. according to $p_{U}$. 
\end{theorem}

\subsection{Achievability proof}
\label{Subsection:Direct_2}
Assume $\left(R_0, R_{1}\right)$ is in the interior of $\mathcal{R}_{\text{2-node}}(\omega)$. We
need to construct a code that consists of an encoding channel $F(i|x^{n},j)
$
and a c-q decoding channel $%\rho_{X^{n}B^{n}}^{U\left(I,J\right)}=
\mathcal{D} _{IJ\to B^n}$, such that the error requirement in \eqref{Equation:Error_2} holds.
% \begin{multline}
% \Big\Vert \omega_{XB}^{\otimes n}-
% \frac{1}{2^{n R_0}}\sum_{j\in [2^{n R_0}]}
% \sum_{x^n\in\mathcal{X}^n} p_X^n(x^n) \ketbra{x^n}_{X^n}
% \\
% \otimes
% %
% \sum_{i\in [2^{nR_1}]} F(i|x^{n},j)
% %
% \mathcal{D}_{IJ\to B^n}(i,j)\Big\Vert _{1}\leq \varepsilon \,.
% \end{multline}
%For simplicity, we omit the dependency of the coding maps on $n$. 

By the definition of $\mathscr{S}_{\text{2-node}}(\omega)$, there exists a c-q
state %$\sigma_{XUB}$ that can be written as
\begin{align}
\sigma_{UXB}=\sum_{%x\in\mathcal{X},
u\in\mathcal{U}}p_{U}\left(u\right)\ketbra{u}_{U}\otimes 
\sigma_{XB}^{u}
\end{align}
such that
\begin{align}
\sigma_{XB}^{u}=&\sum_{x\in\mathcal{X}}p_{X|U}\left(x|u\right)\ketbra{x}_{X}\otimes
\theta_{B}^{u} \,,\; u\in\mathcal{U}
\label{Equation:Sigma_given_U}
\\
\sigma_{XB}  =&\omega_{XB} \,,
\label{Equation:sigma_omega_XB}
\\ %\end{align}
%for $u\in\mathcal{U}$, 
%and
%\begin{align}
R_{1}  \geq I\left(X;U\right)_{\sigma} \,,&\;%\\
R_0+R_1 \geq I\left(XB;U\right)_{\sigma} \,.
\end{align}
%
% \textcolor{blue}{Check if this is necessary,}
% Express $\rho_{XB},\omega_{XB}$ in the following way
% \begin{align}
% \omega_{XB} & =\sum_{x}p_{X}\left(x\right)\ketbra{x}_{X}\otimes\omega_{B}^{x}\\
% \rho_{XB} & =\sum_{x}p_{X}\left(x\right)\ketbra{x}_{X}\otimes\rho_{B}^{x}\\
% \rho_{XB} & =\omega_{XB}\Rightarrow\rho_{B}^{x}=\omega_{B}^{x}\;\;\;\forall x\in\mathcal{X}
% \end{align}
%
% \textcolor{blue}{Check if this is necessary,}
% Define:
% \begin{align}
% p_{X,U}^{*}\left(x,u\right) & =p_{U}\left(u\right)p_{X|U}^{*}\left(x|u\right)
% \end{align}
% Conditioning on $U=u$,  denote
% \begin{align}
% \sigma_{XB}^{u}=\sum_{x\in\mathcal{X}}p_{X|U}\left(x|u\right)\ketbra{x}_{X}\otimes
% \theta_{B}^{u}
% \label{Equation:Sigma_given_U}
% \end{align}
% for $u\in\mathcal{U}$.

\textbf{Classical codebook generation:}
Select a random codebook $\mathscr{C}=\{ u^n(i,j) \}$ by drawing $2^{n(R_0+R_1)}$
i.i.d. sequences according to the distribution $p_{U}^{n}\left(u^{n}\right)=\prod_{k=1}^{n}p_{U}\left(u_{k}\right)$.
Reveal the codebook to Alice and Bob. 

Let $(i,j)$ be a pair of random indices, uniformly distributed
over $
[2^{n R_{1}}]\times [2^{n  R_0}]
$.
Define the following PMF
\begin{align}
\widetilde{P}_{X^{n}IJ}(x^{n},i,j) 
% & =p_{IJ}(i,j) \prod_{k=1}^{n}p_{X|U}\left(x_{k}|u_k\left(i,j\right)\right)
% \nonumber\\
 & \equiv\frac{1}{2^{n(R_0+R_1)}} p_{X|U}^{n}\left(x^n|u^n\left(i,j\right)\right) \,.
 \label{Equation:tilde_P}
\end{align}

\textbf{Encoder:}
We define the encoding channel $F$ 
as the conditional distribution
above, i.e.,
%
%$\widetilde{P}_{I|X^{n},J}$. That is, 
$F%(i|x^n,j)
=\widetilde{P}_{I|X^{n}J}%(i|x^{n},j)
$.

\textbf{Decoder:}
As Bob receives $i$ from Alice, and the random element $j$, he prepares the output state
$%\begin{align}
\mathcal{D}_{IJ\to B^n}(i,j) %& 
=\theta_{B^{n}}^{u^n(i,j)} %\,.
$. %\end{align}

\textbf{Error analysis:}
Let $\delta>0$.
The encoder sends $i\sim F(\cdot|x^{n},j)$.
%At first, consider a fixed $j\in [2^{n R_0}]$. 
%
Given $J=j$, by the classical resolvability theorem,
Cuff
\cite{cuff2008communication} has shown that 
$R_{0}\geq I\left(X;U\right)_{\sigma}$ guarantees
%applying  according to 
 %we get
\begin{align}
\mathbb{E}\left\Vert \widetilde{P}_{JX^{n}}-p_{J}\times p_{X}^n \right\Vert _{1}\leq \delta
 \label{Equation:Resolvability_JX}
\end{align} 
for sufficiently large $n$, where
$\widetilde{P}_{JX^{n}}$ is as in
\eqref{Equation:tilde_P}.
Recall that $\widetilde{P}_{JX^{n}}$ is random, since  the  codebook $\mathscr{C}$ is random. Hence, the expectation is over all realizations of $\mathscr{C}$.
%
% \begin{align}
% Q_{X^n|I,J}(x^n|i,j)= p_{X|U}^n(x^n|u^n(i,j)) \,.
% \end{align}
%
%For a given codebook $\mathscr{C}=\{ u^n(i,j) \}$,
The resulting state is
%
% Applying the decoder to the average value of $I$ when $J=j$ is known
% to him yields:
\begin{align}
\widehat{\rho}_{X^n B^n}
=\frac{1}{2^{n R_0}}\sum_{j,x^n}
\Big( p_X^n(x^n)  \ketbra{x^n}_{X^n}
% \nonumber\\
% &
\otimes  
\sum_{i\in [2^{nR_1}]} \widetilde{P}_{I|X^nJ}(i|x^n,j) \theta_{B^{n}}^{u^n(i,j)} \Big)
\end{align}

According to 
\eqref{Equation:Resolvability_JX}, the probability distributions $\widetilde{P}_{J,X^n}$ and $p_{J}\times p_X^n$ are close on average. Then, let
\begin{align}
\widehat{\tau}_{X^n B^n}
\equiv
\sum_{j,x^n}
\widetilde{P}_{J X^n}(j x^n)  \ketbra{x^n}_{X^n}
% \nonumber\\
% &
\otimes  
\sum_{i\in [2^{nR_1}]} \widetilde{P}_{I|X^nJ}(i|x^n,j) 
\theta_{B^{n}}^{u^n(i,j)} \,.
\end{align}
By \eqref{Equation:Resolvability_JX},
it follows that
\begin{align}
\mathbb{E}\left\Vert \widehat{\tau}_{X^n B^n}-\widehat{\rho}_{X^n B^n}\right\Vert _{1}\leq \delta \,.
 \label{Equation:Sigma_Rho_Distance}
\end{align}

Observe that
\begin{align}
\widehat{\tau}_{X^n B^n}
&=
\sum_{i,j,x^n}
\widetilde{P}_{IJX^n}(i,j,x^n)  \ketbra{x^n}_{X^n}
\otimes  
\theta_{B^{n}}^{u^{n}(i,j)} 
\nonumber\\
&=
\frac{1}{2^{n(R_0+R_1)}}\sum_{i,j,x^n}
p_{X|U}^n(x^n|u^n(i,j))  \ketbra{x^n}_{X^n}
\otimes  
\theta_{B^{n}}^{u^{n}(i,j)} 
\nonumber\\
&=
\frac{1}{2^{n(R_0+R_1)}}\sum_{i,j} 
\sigma_{X^n B^{n}}^{u^{n}(i,j)} 
\end{align}
%
% calculating the average state over all values of $j\in\mathcal{J}$
% yields
%
% Applying this decoding channel on the uniform mix of all the pairs
% $\left(I,J\right)$ gives
%
% \[
% \mathcal{D}_{n}\left(\frac{1}{2^{n(R_{1}+ R_0)}}\sum_{i,j}\ketbra{I}\otimes\ketbra{J}\right)=\frac{1}{2^{n(R_{1}+ R_0)}}\sum_{i,j}\bar{\rho}_{B^{n}}^{u^{n}\left(i,j\right)}
% \]
where the second equality is due to the definition of $\widetilde{P}$ in \eqref{Equation:tilde_P}, and the last line follows from \eqref{Equation:Sigma_given_U}.

Thus, according to the quantum resolvability theorem,
Theorem~\ref{Theorem:Quantum_resolvability}, when applied to the joint system $XB$,
we have
\begin{align}
&\mathbb{E}\norm{ \sigma_{XB}^{{\otimes n}}-
\widehat{\tau}_{X^n B^n}}_1 \leq \delta
\label{Equation:Resolvability_XB}
\end{align}
for sufficiently large $n$.
Therefore, by the triangle inequality,
\begin{align}
\mathbb{E}\norm{ \omega_{XB}^{{\otimes n}}-
\widehat{\rho}_{X^n B^n}}_1
% \nonumber\\
% &
&\leq 
\mathbb{E}\norm{ \omega_{XB}^{{\otimes n}}-
\widehat{\tau}_{X^n B^n}}_1
+ 
% \nonumber\\
% &
\mathbb{E}\norm{ 
\widehat{\tau}_{X^n B^n}-
\widehat{\rho}_{X^n B^n}}_1
\nonumber\\
&\leq 2\delta
\end{align}
by \eqref{Equation:sigma_omega_XB}, \eqref{Equation:Sigma_Rho_Distance} and \eqref{Equation:Resolvability_XB}. 
\qed

\subsection{Converse proof}
Let $\left(R_0,R_1\right)$ be an achievable rate pair. Then, there
exists a sequence $\left(2^{nR_{0}},2^{n R_1},n\right)$ coordination codes such
that the joint quantum state $\widehat{\rho}_{X^{n}B^{n}}$ satisfies
\begin{align}
\left\Vert \omega_{XB}^{\otimes n}-\widehat{\rho}_{X^{n}B^{n}}\right\Vert _{1} & \leq \varepsilon_n
\label{Equation:Converse_Error}
\end{align}
where $\varepsilon_n$ tends to zero as $n\to\infty$.

Fix an index $k\in\left\{ 1,\dots,n\right\} $.
%$K$ will serve as a random time index in the proof. 
By trace monotonicity \cite{wilde2017quantum}, taking the
% We now prove that the state $\widehat{\rho}_{X^{n}B^{n}}$ satisfies
%
% \[
% \left\Vert \widehat{\rho}_{X^{n}B^{n}}-\bigotimes_{k=1}^{n}\widehat{\rho}_{X_{k}B_{k}}\right\Vert _{1}<\varepsilon
% \]
%
partial trace over $X_j,B_j$, $j\neq k$, maintains the inequality. 
Thus,
\begin{align}
\left\Vert \omega_{X B }-\widehat{\rho}_{X_{k}B_{k}}\right\Vert _{1} & \leq\varepsilon_n \,.
\label{Equation:Error_k_TwoNodes}
\end{align}
%for $k\in [n]$.
% By the Fuchs–van de Graaf inequalities,
% \begin{align}
% 1-\sqrt{F(\rho,\sigma)}\leq \frac{1}{2} \norm{\rho-\sigma}_1\leq 
% \sqrt{1-F(\rho,\sigma)} \,.
% \end{align}
% Based on the lower bound,
% \begin{align}
% F(\rho,\sigma)\geq \left( 1-\frac{1}{2} \norm{\rho-\sigma}_1 \right)^2
% \end{align}
% Thus,
% \begin{align}
% F\left(\bigotimes_{k=1}^n\widehat{\rho}_{X_{k}B_{k}},\omega_{XB}^{\otimes n}\right)
% &=
% \prod_{k=1}^n F\left(\widehat{\rho}_{X_{k}B_{k}},\omega_{XB}\right)
% \nonumber\\
% &\geq \prod_{k=1}^n\left( 1-\frac{1}{2} \norm{\widehat{\rho}_{X_{k}B_{k}}-\omega_{XB}}_1 \right)^{2}
% \end{align}
%
Then,
by the AFW inequality \cite{AFW_Winter_2016}, % and entropy subadditivity, 
% and the entropy chain rule, and since conditioning
% cannot increase entropy we get
\begin{align}
\abs{H\left(X^{n}B^{n}\right)_{\widehat{\rho}}-n H\left(X B \right)_{\omega}} & \leq n\beta_n \,,
\label{Equation:AFW_Entropy_TwoNodes_1}
\intertext{and}
\abs{H\left(X_k B_k\right)_{\widehat{\rho}}- H\left(X B \right)_{\omega}} & \leq \beta_n \,,
\label{Equation:AFW_Entropy_TwoNodes_2}
%\\
% \left|\sum_{k=1}^{n}H\left(X_{k},B_{k}\right)_{\widehat{\rho}_{X_{k}B_{k}}}\right.\\
% -\left.\sum_{k=1}^{n}H\left(X_{k},B_{k}|K=k\right)_{\widehat{\rho}_{X_{k}B_{k}}}\right| & \leq ng(\varepsilon)\\
% \iff I\left(X_{k},B_{K};K\right) & \leq ng(\varepsilon)
\end{align}
for $k\in [n]$,
where $\beta_n$ tends to zero as $n\to \infty$.
Therefore,
% Then,
% by the AFW inequality \cite{AFW_Winter_2016}, % and entropy subadditivity, 
% % and the entropy chain rule, and since conditioning
% % cannot increase entropy we get
\begin{align}
\abs{H\left(X^{n}B^{n}\right)_{\widehat{\rho}}-\sum_{k=1}^{n}H\left(X_{k}B_{k}\right)_{\widehat{\rho}}}
&\leq 
\abs{H\left(X^{n}B^{n}\right)_{\widehat{\rho}}-n H\left(X B \right)_{\omega}}
+
\abs{n H\left(X B \right)_{\omega}-\sum_{k=1}^{n}H\left(X_{k}B_{k}\right)_{\widehat{\rho}}}
\nonumber\\
&\leq 
\abs{H\left(X^{n}B^{n}\right)_{\widehat{\rho}}-n H\left(X B \right)_{\omega}}
+\sum_{k=1}^{n}
\abs{ H\left(X B \right)_{\omega}-H\left(X_{k}B_{k}\right)_{\widehat{\rho}}}
\nonumber\\
& \leq 2n\beta_n
\,.
\label{Equation:AFW_Entropy}
%\\
% \left|\sum_{k=1}^{n}H\left(X_{k},B_{k}\right)_{\widehat{\rho}_{X_{k}B_{k}}}\right.\\
% -\left.\sum_{k=1}^{n}H\left(X_{k},B_{k}|K=k\right)_{\widehat{\rho}_{X_{k}B_{k}}}\right| & \leq ng(\varepsilon)\\
% \iff I\left(X_{k},B_{K};K\right) & \leq ng(\varepsilon)
\end{align}
% for $k\in [n]$,
% where $\beta_n$ tends to zero as $n\to \infty$.

Now, we have
\begin{align}
% nR_1 &\geq H(I)
% \\
% &\geq H(I|J)
% \\
% &\geq I(X^n;I|J)
% \\
% &= I(X^n;I,J) \,,
% \\
n(R_0+R_1)&\geq 
H(IJ)
\\
&\geq I(X^nB^n;IJ)_{\widehat{\rho}} 
\end{align}
since % conditioning cannot increase entropy,  
the conditional entropy is nonnegative for classical and c-q states, and the CR element $J$ is statistically independent of the source $X^n$.
Furthermore, by  entropy 
sub-additivity \cite{wilde2017quantum},
\begin{align}
I(X^nB^n;IJ)_{\widehat{\rho}} 
&\geq 
H(X^nB^n)_{\widehat{\rho}} -\sum_{k=1}^n H(X_k B_k|IJ)_{\widehat{\rho}} 
\nonumber\\
&\geq 
\sum_{k=1}^n I(X_kB_k;IJ)_{\widehat{\rho}} -2n\beta_n
\end{align}
where the last inequality follows from 
\eqref{Equation:AFW_Entropy}.
Defining a time-sharing variable
$K\sim \mathrm{Unif}[n]$, this can be written as
\begin{align}
R_0+R_1+2\beta_n \geq  I(X_K B_K;I J|K)_{\widehat{\rho}}
\label{Equation:Information_given_K}
\end{align}
with respect to the extended state:
\begin{align}
\widehat{\rho}_{KIJX_K B_K}=\frac{1}{n}\sum_{k=1}^n \ketbra{k}\otimes \widehat{\rho}_{IJX_k B_k}
\,.
\end{align}

Observe that by \eqref{Equation:Error_k_TwoNodes} and  the triangle inequality,
\begin{align}
\left\Vert \omega_{X B }-\widehat{\rho}_{X_{K}B_{K}}\right\Vert _{1}
&=\left\Vert \omega_{X B }-\frac{1}{n}\sum_{k=1}^n\widehat{\rho}_{X_{k}B_{k}}\right\Vert _{1} 
\nonumber\\
& 
\leq\varepsilon_n  \,.
\end{align}
Thus, by the AFW inequality, 
\begin{align}
I(X_K B_K; K)_{\widehat{\rho}}
&=
H(X_K B_K)_{\widehat{\rho}}-\frac{1}{n}
\sum_{k=1}^n H(X_k B_k)_{\widehat{\rho}} 
\nonumber\\
&
\leq \gamma_n\,,
\end{align} where $\gamma_n$ tends to zero. % as $n\to \infty$.
Together with \eqref{Equation:Information_given_K}, it follows that 
\begin{align}
R_0+R_1+2\beta_n+\gamma_n \geq  I(X_K B_K;I J K)_{\widehat{\rho}}
\end{align}
By similar arguments,
% \begin{align}
%  nR_1 &\geq H(I)
%  \\
%  &\geq H(I|J)
%  \\
%  &\geq I(X^n;I|J)
%  \\
%  &= I(X^n;I,J)
% \end{align}
\begin{align}
 R_1+2\beta_n+\gamma_n &\geq 
  I(X_K;I J)
\end{align}
To complete the converse proof, we identify $U$, $X$, and $B$ with 
$(I,J,K)$, $X_K$, and $B_K$, respectively.
Observe that given $(i,j,k)$, the joint state of $X_K$ and $B_K$ is 
$%\begin{align}
\left(\sum_{x_k\in\mathcal{X}} p_{X_k|IJ}(x_k|i,j) \ketbra{x_k}_{X_K}\right)
\otimes  \rho_{B_k}^{(i,j)} 
$, %\end{align}
where $p_{X^n|IJ}$ is the a posteriori probability distribution.
Thus, there $X$ and $B$ are uncorrelated when conditioned on $U$, as required.

The bound on $|\mathcal{U}|$ follows by applying the Caratheodory theorem to the real-valued parameteric representation of density matrices, as in 
\cite[App. B]{Pereg:21p}. \qed

\section{No-Communication Network Analysis}
Consider  the no-communication network in Figure~\ref{Figure 2 - No-communication network}, of a quantum state $\omega_{ABC}$.
To prove %the capacity theorem, 
Theorem~\ref{Theorem:NC}, we use similar tools. % as in Section~\ref{Subsection:Proof_2}.
%The achievability part of the proof in this subsection uses the quantum resolvability result presented in theorem \eqref{Equation:Quantum_resolvability}
%
The achievability proof is straightforward, and it is thus omitted. 
Then, consider the converse part.
%\subsection{Converse proof}
Assume that 
$R_0$ is achievable. Therefore, there exists a sequence of $(2^{nR_0},n)$ of coordination codes such that for sufficiently large values of $n$,
\begin{align} \norm{\widehat{\rho}_{A^n B^n C^n}-\omega_{A B C}^{\otimes n}}_1 \leq \varepsilon _n \,,
\end{align}
where $\varepsilon_n\to0$ as $n\to\infty$.
%We need to prove that $R_0\geq \mathsf{C}_{\text{NC}}(\omega)$. 

Applying the chain rule, % for entropy and the above consideration, we get
\begin{align}
    nR_0&\geq H(J)
    \\
    &\geq I(A^nB^nC^n;J)_{\widehat{\rho}}
    % \\
    % &= I(A^nB^nC^n;U)_{\widehat{\rho}}
    \\
    &=\sum_{k=1}^n I(A_kB_kC_k;J|A^{k-1}B^{k-1}C^{k-1})_{\widehat{\rho}}
\label{Equation:Converse_NC1}
\end{align}

% \textcolor{red}{According to similar considerations leading to} \eqref{Equation:IID_2}, we have 

For every $k\in[n] $, by trace monotonicity \cite{wilde2017quantum}, %taking the
% We now prove that the state $\widehat{\rho}_{X^{n}B^{n}}$ satisfies
%
% \[
% \left\Vert \widehat{\rho}_{X^{n}B^{n}}-\bigotimes_{k=1}^{n}\widehat{\rho}_{X_{k}B_{k}}\right\Vert _{1}<\varepsilon
% \]
%
% partial trace over $A_j,B_j,C_j$, $j> k$, maintains the inequality. 
% Thus,
\begin{align}
\left\Vert \omega_{A B C}^{\otimes k}-\widehat{\rho}_{A^{k}B^{k}C^k}\right\Vert _{1} & \leq\varepsilon_n \,.
% \label{Equation:Error_k_No_Comm}
\end{align}
Then,
by the AFW inequality \cite{AFW_Winter_2016} \cite[Ex. 11.10.2]{wilde2017quantum}, % and entropy subadditivity, 
% and the entropy chain rule, and since conditioning
% cannot increase entropy we get
\begin{align}
% \abs{H\left(X^{n}B^{n}\right)_{\widehat{\rho}}-n H\left(X B \right)_{\omega}} & \leq n\beta_n \,,
% \label{Equation:AFW_Entropy_1}
% \intertext{and}
\abs{I(A_kB_kC_k;A^{k-1}B^{k-1}C^{k-1})_{\widehat{\rho}}- I(A_kB_kC_k;A^{k-1}B^{k-1}C^{k-1})_{\omega^{\otimes k}}}%_{\omega\otimes [\omega^{\otimes (k-1)}]} }
& \leq \beta_n \,,
\label{Equation:AFW_MI_2}
%\\
% \left|\sum_{k=1}^{n}H\left(X_{k},B_{k}\right)_{\widehat{\rho}_{X_{k}B_{k}}}\right.\\
% -\left.\sum_{k=1}^{n}H\left(X_{k},B_{k}|K=k\right)_{\widehat{\rho}_{X_{k}B_{k}}}\right| & \leq ng(\varepsilon)\\
% \iff I\left(X_{k},B_{K};K\right) & \leq ng(\varepsilon)
\end{align}
where $\beta_n$ tends to zero as $n\to \infty$. That is,
\begin{align}
% \abs{H\left(X^{n}B^{n}\right)_{\widehat{\rho}}-n H\left(X B \right)_{\omega}} & \leq n\beta_n \,,
% \label{Equation:AFW_Entropy_1}
% \intertext{and}
I(A_kB_kC_k;A^{k-1}B^{k-1}C^{k-1})_{\widehat{\rho}} & \leq \beta_n
\label{Equation:AFW_MI_3}
\end{align}
since $A_k B_k C_k$ and
$(A_j B_j C_j)_{j<k}$ are in a product state
$\omega\otimes\omega^{\otimes (k-1)}$.
% \begin{align}
% \left\Vert \widehat{\rho}_{A^{n} B^{n} C^{n}}-\bigotimes_{k=1}^{n}\widehat{\rho}_{A_{k} B_{k} C_{k}}\right\Vert _{1}\leq 2 \varepsilon_n\;.
% \end{align} 
% Therefore, by the AFW inequality \cite{AFW_Winter_2016},  %there exists $\delta_n$ that tends to zero as $n \to \infty$, such that 
% \begin{align}
% I(A_k B_k C_k;A^{k-1} B^{k-1} C^{k-1})_{\widehat{\rho}} 
% &\leq \beta_n \,.
% \end{align}
Hence, by \eqref{Equation:Converse_NC1},
\begin{align}
    nR_0%&\geq \sum_{k=1}^n I(A_k B_k C_k;U|A^{k-1} B^{k-1} C^{k-1}) \\ 
    &\geq\sum_{k=1}^nI(A_k B_k C_k;J A^{k-1} B^{k-1}C^{k-1})_{\widehat{\rho}}-n\beta_n \nonumber\\ 
    &\geq\sum_{k=1}^nI(A_k B_k C_k;J)_{\widehat{\rho}}-n\beta_n \nonumber\\
    &\geq n\left( \inf_{\sigma_{UABC}\in\mathscr{S}_{NC}(\omega)}
%\exists\rho_{XUB}\in \mathscr{S}\;s.t.\\
 I(U;ABC)_\sigma-2\beta_n \right)
\end{align}
taking $U\equiv J$, as
%Notice that given $J$, 
the encoders are uncorrelated given $J$. % (they only depend on $j$ for the state preparation). 
%Thus, we can identify the auxiliary variable $U$ as $J$. 
\qed

\section{Broadcast Network Analysis}
Consider coordination in broadcast network, as in Figure~\ref{Figure 3: Broadcast_network} in the main text, of a classical-quantum-quantum state $\omega_{XB_1 B_2}$.
To prove the capacity theorem, Theorem~\ref{Theorem: Broadcast}, we use similar tools as in Section~\ref{Subsection:Proof_2}.

\subsection{Achievability proof}
Assume $\left(R_{0},R_{1}\right)$ is in the interior of $\mathcal{R}_{\text{BC}}(\omega)$. We need to construct a code that consists of an encoding channel $F(i|x^{n},j)$ and a two c-q decoding channels $\mathcal{D}_{IJ\to B_{1}^{n}}$ and $\mathcal{D}_{IJ\to B_{2}^{n}}$,such that 
\begin{align}
    \Big\Vert\omega_{XB}^{\otimes n}-\frac{1}{2^{nR_{0}}}\sum_{j\in[2^{nR_{0}}]}\sum_{x^{n}\in\mathcal{X}^{n}}p_{X}^{n}(x^{n})\ketbra{x^{n}}_{X^{n}}\otimes
    % \nonumber\\
    %\otimes
\sum_{i\in[2^{nR_{1}}]}F(i|x^{n},j)\mathcal{D}_{IJ\to B_{1}^{n}}(i,j)\otimes \mathcal{D}_{IJ\to B_{2}^{n}}(i,j)\Big\Vert_{1}\leq\varepsilon \,.
\end{align}
According to the definition of $\mathscr{S}_{\text{BC}}(\omega)$ (see Subsection~\ref{Subsection:BC}), there exists a c-q state $\sigma_{XUB_{1}B_{2}} $ that can be written as
\begin{align}
\sigma_{XUB_{1}B_{2}}=\sum\limits _{(x,u)\in\mathcal{X}\times\mathcal{U}}p_{X,U}(x,u)\ketbra{x}_{X}\otimes\ketbra{u}_{U}\otimes\theta_{B_{1}}^{u}\otimes\eta_{B_{2}}^{u}
\end{align} and satisfy \begin{align}
\sigma_{XB_{1}B_{2}}=\omega_{XB_{1}B_{2}}
\label{Equation:sigma_omega_XB1B2}
\end{align} We will also consider conditioning on $U=u$, and denote 
\begin{align}
\sigma_{XB_{1}B_{2}}^{u}=\sum_{x\in\mathcal{X}}p_{X|U}(x|u)\ketbra{x}_{X}\otimes\theta_{B_{1}}^{u}\otimes\eta_{B_{2}}^{u} \,.
\label{Equation:Sigma_given_U_B}
\end{align}

\textbf{Classical codebook generation:} 
Select a random codebook $\mathscr{C}_{\text{BC}}=\{u^{n}(i,j)\}$ by drawing $2^{n(R_{0}+R_{1})}$ i.i.d. sequences according to the distribution $p_{U}^{n}%\left(u^{n}\right)=\prod_{k=1}^{n}p_{U}\left(u_{k}\right)
$. Reveal the codebook. % to Alice and Bob. 

% Let $\widetilde{P}_{X^{n}IJ}$ be a joint distribution as in  \eqref{Equation:tilde_P}.
% $(i,j)$ be a pair of random indices, uniformly distributed over $[2^{nR_{1}}]\times[2^{nR_{0}}]$. Define the following PMF
% \begin{align}
%     &\widetilde{P}_{X^{n}IJ}(x^{n},i,j)=
%     \nonumber\\
%     &=p_{IJ}(i,j)\prod_{k=1}^{n}p_{X|U}\left(x_{k}|U_{k}\left(i,j\right)\right)\nonumber\\
%     &\equiv\frac{1}{2^{n(R_{0}+R_{1})}}p_{X|U}^{n}\left(x^{n}|u^{n}\left(i,j\right)\right)\\
%     \label{Equation:tilde_{P}}
% \end{align}

\textbf{Encoder:} 
Define the encoding channel  as $ \ensuremath{F%(i|x^{n},j)
=\widetilde{P}_{I|X^{n}J}%(i|x^{n},j)
}$, where  $\widetilde{P}_{X^{n}IJ}$ be a joint distribution as in  \eqref{Equation:tilde_P}.

\textbf{Decoders:}
As Bob~1 and Bob~2 receive $i$ from Alice, and the random element $j$, they prepare the following output states, 
\begin{align}
    \mathcal{D}^{(1)}_{IJ\to B_{1}^{n}}(i,j)&=\theta_{B_{1}}^{u^{n}(i,j)} \,,
    \\
    \mathcal{D}^{(2)}_{IJ\to B_{2}^{n}}(i,j)&=\eta_{B_{2}}^{u^{n}(i,j)} \,.
\end{align}

\textbf{Error analysis:}
Let $\delta>0$. The encoder sends $i\sim F(\cdot|x^{n},j)$. 
As in Subsection~\ref{Subsection:Direct_2},
given $j$, if %by the classical resolvability theorem, Cuff \cite{cuff2008communication} has shown that 
$R_{1}\geq I\left(X;U\right)$, then 
\begin{align}
    \mathbb{E}\left\Vert \widetilde{P}_{JX^{n}}-p_{J}\times p_{X}^{n}\right\Vert _{1}\leq\delta
\label{Equation:Resolvability_JX_B}
\end{align}
for sufficiently large $n$. %, where $\widetilde{P}_{JX^{n}}$ is as in \eqref{Equation:tilde_P}. 
As $\widetilde{P}_{JX^{n}}$  depends on the random codebook $\mathscr{C}_{\text{BC}}$, the expectation is over all realizations of $\mathscr{C}_{\text{BC}}$. The resulting state is
\begin{align}
%&
\widehat{\rho}_{X^{n}B_1^{n}B_2^{n}}
%\nonumber\\
&=\frac{1}{2^{nR_{0}}}\sum_{j\in[2^{nR_{0}}]}\sum_{x^{n}\in\mathcal{X}^{n}}\Big(p_{X}^{n}(x^{n})\ketbra{x^{n}}_{X^{n}}
% \nonumber 
% \\
% &
\otimes\sum_{i\in[2^{nR_{1}}]}F(i|x^{n},j)\mathcal{D}_{IJ\to B_{1}^{n}}(i,j)\otimes\mathcal{D}_{IJ\to B_{2}^{n}}(i,j)\Big)
\nonumber\\
&=\frac{1}{2^{nR_{0}}}\sum_{j,x^{n}}\Big(p_{X}^{n}(x^{n})\ketbra{x^{n}}_{X^{n}}
% \nonumber
% \\
% &
\otimes\sum_{i\in[2^{nR_{1}}]}\widetilde{P}_{I|X^{n}J}(i|x^{n},j)\theta_{B_{1}}^{u^{n}(i,j)}\otimes\eta_{B_{2}}^{u^{n}(i,j)}\Big) \,.
\end{align}
According to \eqref{Equation:Resolvability_JX_B}, the probability distributions $\widetilde{P}_{JX^{n}}$ and $p_{J}\times p_{X}^{n}$ are close on average. Then, let
\begin{align}
\widehat{\tau}_{X^{n}B_{1}^{n}B_{2}^{n}}\equiv\sum_{j,x^{n}}\widetilde{P}_{JX^{n}}(j,x^{n})\ketbra{x^{n}}_{X^{n}}
% \nonumber
% \\
% &
\otimes\sum_{i\in[2^{nR_{1}}]}\widetilde{P}_{I|X^{n}J}(i|x^{n},j)\theta_{B_{1}}^{u^{n}(i,j)}\otimes\eta_{B_{2}}^{u^{n}(i,j)} \,.
\end{align}
Then, it follows that
\begin{align}
    \mathbb{E}\left\Vert \widehat{\tau}_{X^{n}B_{1}^{n}B_{2}^{n}}-\widehat{\rho}_{X^{n}B_{1}^{n}B_{2}^{n}}\right\Vert _{1}\leq\delta \,,
    \label{Equation: Tau_rho}
\end{align}
by \eqref{Equation:Resolvability_JX_B}.
Observe that
\begin{align}
\widehat{\tau}_{X^{n}B_{1}^{n}B_{2}^{n}}
&=\sum_{i,j,x^{n}}\widetilde{P}_{IJX^{n}}(i,j,x^{n})\ketbra{x^{n}}_{X^{n}}
% \nonumber\\
% &
\otimes\theta_{B_{1}}^{u^{n}(i,j)}\otimes\eta_{B_{2}}^{u^{n}(i,j)}\nonumber\\
&=\frac{1}{2^{n(R_{0}+R_{1})}}\sum_{i,j,x^{n}}p_{X|U}^{n}(x^{n}|u^{n}(i,j))\ketbra{x^{n}}_{X^{n}}
% \nonumber\\
% &
\otimes\theta_{B_{1}}^{u^{n}(i,j)}\otimes\eta_{B_{2}}^{u^{n}(i,j)}
\nonumber\\
&=\frac{1}{2^{n(R_{0}+R_{1})}}\sum_{i,j}\sigma_{X^{n}B_{1}^{n}B_{2}^{n}}^{u^{n}(i,j)} \,,
\end{align}
where the second equality is due to the definition of $\widetilde{P}$ in \eqref{Equation:tilde_P}, and the last line follows from \eqref{Equation:Sigma_given_U_B}.

Thus, according to the quantum resolvability theorem \ref{Theorem:Quantum_resolvability}, when applied to the joint system $XB_{1}B_{2}$, we have 
\begin{align}
\mathbb{E}\norm{\sigma_{XB_{1}^{n}B_{2}^{n}}^{{\otimes n}}-\widehat{\tau}_{X^{n}B_{1}^{n}B_{2}^{n}}}_{1}\leq\delta
\label{Equation:Tau_sigma}
\end{align}
for sufficiently large $n$. Therefore, by the triangle inequality,
\begin{align}\mathbb{E}\norm{\omega_{XB_{1}B_{2}}^{{\otimes n}}-\widehat{\rho}_{X^{n}B_{1}^{n}B_{2}^{n}}}_{1}\nonumber 
&\leq\mathbb{E}\norm{\omega_{XB_{1}^{n}B_{2}^{n}}^{{\otimes n}}-\widehat{\tau}_{X^{n}B_{1}^{n}B_{2}^{n}}}_{1}+\mathbb{E}\norm{\widehat{\tau}_{X^{n}B_{1}^{n}B_{2}^{n}}-\widehat{\rho}_{X^{n}B_{1}^{n}B_{2}^{n}}}_{1}\nonumber
\\
&\leq2\delta
\end{align}
by \eqref{Equation:sigma_omega_XB1B2}, \eqref{Equation: Tau_rho} and \eqref{Equation:Tau_sigma}. \qed

\subsection{Converse proof}
Let $\left(R_{0},R_{1}\right)$ be an achievable coordination rate pair for the simulation of a c-q-q state $\omega_{XB_1 B_2}$ in the broadcast setting. Then, there exists a sequence of $\left(2^{nR_{0}},2^{nR_{1}},n\right)$ coordination codes such that the joint quantum state $\widehat{\rho}_{X^{n}B_{1}^{n}B_{2}^{n}}$ satisfies 
\begin{align}
    \left\Vert \omega_{XB_{1}^{n}B_{2}^{n}}^{\otimes n}-\widehat{\rho}_{X^{n}B_{1}^{n}B_{2}^{n}}\right\Vert _{1}\leq\varepsilon_{n}\,,
\label{Equation:Converse_Error_Broadcast}
\end{align} 
%%%%%%%%%%%%%%%%
where $\varepsilon_{n}$ tends to zero as $n\to\infty$.
Fix an index $k\in\left\{ 1,\dots,n\right\}$. By trace monotonicity \cite{wilde2017quantum}, taking the partial trace over $X_{j}$, $B_{1j}$, $B_{2j}$, for $j\neq k$, maintains the inequality, thus
% We now prove that the state $\widehat{\rho}_{X^{n}B^{n}}$ satisfies
%
% \[
% \left\Vert \widehat{\rho}_{X^{n}B^{n}}-\bigotimes_{k=1}^{n}\widehat{\rho}_{X_{k}B_{k}}\right\Vert _{1}<\varepsilon
% \]
%
Thus,
\begin{align}
\left\Vert \omega_{X B_1 B_2 }-\widehat{\rho}_{X_{k}B_{1k}B_{2k}}\right\Vert _{1} & \leq\varepsilon_n \,.
\label{Equation:Error_k_Broadcast}
\end{align}
%for $k\in [n]$.
% By the Fuchs–van de Graaf inequalities,
% \begin{align}
% 1-\sqrt{F(\rho,\sigma)}\leq \frac{1}{2} \norm{\rho-\sigma}_1\leq 
% \sqrt{1-F(\rho,\sigma)} \,.
% \end{align}
% Based on the lower bound,
% \begin{align}
% F(\rho,\sigma)\geq \left( 1-\frac{1}{2} \norm{\rho-\sigma}_1 \right)^2
% \end{align}
% Thus,
% \begin{align}
% F\left(\bigotimes_{k=1}^n\widehat{\rho}_{X_{k}B_{k}},\omega_{XB}^{\otimes n}\right)
% &=
% \prod_{k=1}^n F\left(\widehat{\rho}_{X_{k}B_{k}},\omega_{XB}\right)
% \nonumber\\
% &\geq \prod_{k=1}^n\left( 1-\frac{1}{2} \norm{\widehat{\rho}_{X_{k}B_{k}}-\omega_{XB}}_1 \right)^{2}
% \end{align}
%
Then,
by the AFW inequality \cite{AFW_Winter_2016}, % and entropy subadditivity, 
% and the entropy chain rule, and since conditioning
% cannot increase entropy we get
\begin{align}
\abs{H\left(X^{n}B_1^{n}B_2^{n}\right)_{\widehat{\rho}}-n H\left(X B_1 B_2\right)_{\omega}} & \leq n\beta_n \,,
\label{Equation:AFW_Entropy_Broadcast_1}
\intertext{and}
\abs{H\left(X_k B_{1k}B_{2k}\right)_{\widehat{\rho}}- H\left(X B_1 B_2 \right)_{\omega}} & \leq \beta_n \,,
\label{Equation:AFW_Entropy_Broadcast_2}
%\\
% \left|\sum_{k=1}^{n}H\left(X_{k},B_{k}\right)_{\widehat{\rho}_{X_{k}B_{k}}}\right.\\
% -\left.\sum_{k=1}^{n}H\left(X_{k},B_{k}|K=k\right)_{\widehat{\rho}_{X_{k}B_{k}}}\right| & \leq ng(\varepsilon)\\
% \iff I\left(X_{k},B_{K};K\right) & \leq ng(\varepsilon)
\end{align}
for $k\in [n]$,
where $\beta_n$ tends to zero as $n\to \infty$.
Therefore,
% \begin{align}
%   &\left\Vert \widehat{\rho}_{X^{n}B^{n}}-\bigotimes_{k=1}^{n}\widehat{\rho}_{X_{k}B_{k}}\right\Vert _{1}
%  \nonumber\\
%  % & =\left\Vert \widehat{\rho}_{X^{n}B^{n}}-\omega_{XB}^{\otimes n}+\omega_{XB}^{\otimes n}-\bigotimes_{k=1}^{n}\widehat{\rho}_{X_{k}B_{k}}\right\Vert _{1}\\
%  & \leq\left\Vert \widehat{\rho}_{X^{n}B^{n}}-\omega_{XB}^{\otimes n}\right\Vert _{1}+
%  \left\Vert \omega_{XB}^{\otimes n}-\bigotimes_{k=1}^{n}\widehat{\rho}_{X_{k}B_{k}}\right\Vert _{1}
%  \nonumber\\
%  % & \leq\varepsilon+\prod_{k=1}^{n}\left\Vert \omega_{XB}-\widehat{\rho}_{X^{k}B^{k}}\right\Vert _{1}\\
%  % & \leq\varepsilon+\varepsilon^{n}\\
%  & \leq 2\varepsilon_n
% \label{Equation:IID_2}
% \end{align}
% where the inequalities follow from   the triangle inequality
% and %the second from 
% \eqref{Equation:Converse_Error}-\eqref{Equation:Error_k}, respectively.
% Then,
% by the AFW inequality \cite{AFW_Winter_2016}, % and entropy subadditivity, 
% % and the entropy chain rule, and since conditioning
% % cannot increase entropy we get
\begin{align}
\abs{H\left(X^{n}B_1^{n}B_2^{n}\right)_{\widehat{\rho}}-\sum_{k=1}^{n}H\left(X_{k}B_{1k}B_{2k}\right)_{\widehat{\rho}}} & \leq 2n\beta_n
\,.
\label{Equation:AFW_Entropy_Broadcast}
%\\
% \left|\sum_{k=1}^{n}H\left(X_{k},B_{k}\right)_{\widehat{\rho}_{X_{k}B_{k}}}\right.\\
% -\left.\sum_{k=1}^{n}H\left(X_{k},B_{k}|K=k\right)_{\widehat{\rho}_{X_{k}B_{k}}}\right| & \leq ng(\varepsilon)\\
% \iff I\left(X_{k},B_{K};K\right) & \leq ng(\varepsilon)
\end{align}
%%%%%%%%%%%%%%%%%%%
% According to the AFW inequality \cite{AFW_Winter_2016} and the entropy chain rule, and since  conditioning cannot increase entropy, we have
%
% \begin{align}
% \abs{H\left(X^{n}B_{1}^{n}B_{2}^{n}\right)_{\widehat{\rho}}-\sum_{k=1}^{n}H\left(X_{k}B_{1k}B_{2k}\right)_{\widehat{\rho}}}\leq n\beta_{n}
% \label{Equation:AFW_Entropy_Broadcast}
% \end{align}
%
% for $k\in[n]$, where $\beta_{n}$ tends to zero as $n\to\infty$. 
Now, we also have 
\begin{align}
    n(R_{0}+R_{1})&\geq H(IJ)\nonumber
    \\
    &\geq I(X^{n}B_1^{n}B_2^{n};IJ)_{\widehat{\rho}}\,,
\end{align}
since %conditioning cannot increase entropy, 
the conditional entropy is nonnegative for classical and c-q-q states, and the CR element $J$ is statistically independent of the source $X^{n}$. Furthermore, by entropy sub-additivity \cite{wilde2017quantum},
\begin{align}
I(X^n B_1^n B_2^n;IJ)_{\widehat{\rho}}
%\nonumber\\
&\geq H(X^{n}B_{1}^{n}B_{2}^{n})_{\widehat{\rho}} %\nonumber \\
- \sum_{k=1}^{n}H(X_{k}B_{1k}B_{2k}|IJ)_{\widehat{\rho}}\nonumber
\\
&\geq \sum_{k=1}^{n}I(X_{k}B_{1k}B_{2k};IJ)_{\widehat{\rho}} - 2n\beta_{n}
\label{Equation:Information_given_K_Broadcast}
\end{align} 
where the last inequality follows from \eqref{Equation:AFW_Entropy_Broadcast}.

Defining a time-sharing variable $K\sim\text{Unif}[n]$, this can be written as
\begin{align}
R_0+R_1+2\beta_n \geq I(X_K B_{1K} B_{2K};IJ|K)_{\widehat{\rho}}
\end{align}
with respect to the extended state
\begin{align}
\widehat{\rho}_{KIJX_{k}B_{1k}B_{2k}}=\frac{1}{n}\sum_{k=1}^{n}\ketbra{k}\otimes\widehat{\rho}_{IJX_{k}B_{1k}B_{2k}}\,.
\end{align}
Observe that
\begin{align}
\left\Vert \omega_{X B_1 B_2 }-\widehat{\rho}_{X_{K}B_{1K}B_{2K}}\right\Vert _{1}
=\left\Vert \omega_{X B_1 B_2 }-\frac{1}{n}\sum_{k=1}^n\widehat{\rho}_{X_{k}B_{1k}B_{2k}}\right\Vert _{1} 
 \leq\varepsilon_n 
\end{align}
by the triangle inequality (see \eqref{Equation:Error_k_Broadcast}).
 Thus, by the AFW inequality,
\begin{align}
I(X_{K}B_{1K}B_{2K};K)_{\widehat{\rho}}&=
H(X_{K}B_{1K}B_{2K})_{\widehat{\rho}}-\frac{1}{n}\sum_{k=1}^{n}H(X_{k}B_{1k}B_{2k})_{\widehat{\rho}}
\nonumber\\
&\leq \gamma_n \,,
\end{align}
where $\gamma_n$ tends to zero as $n\to \infty$.
Together with \eqref{Equation:Information_given_K_Broadcast}, it implies
\begin{align}
    R_{0}+R_{1}+2\beta_{n}+\gamma_{n}\geq I(X_{K}B_{1K}B_{2K};IJK)_{\widehat{\rho}}\,.
\end{align}
By similar arguments,
\begin{align}
    R_{1}+2\beta_{n}+\gamma_{n}\geq I(X_{K};IJ)\,.
\end{align}

To complete the converse proof, we identify $U$, $X$, and $B_{1}B_{2}$ with $(I,J,K)$, $X_{K}$, and $B_{1K}B_{2K}$, respectively. Observe that given $(i,j,k)$, the joint state of $X_{K}$ and $B_{1K}B_{2K}$ is
\begin{align}
\left(\sum_{x_{k}\in\mathcal{X}}p_{X_{k}|IJ}(x_{k}|i,j)\ketbra{x_{k}}_{X_{K}}\right)\otimes\rho_{B_{1K}}^{(i,j)}\otimes\rho_{B_{2K}}^{(i,j)}\,,
\end{align}
where $p_{X^{n}|IJ}$ is the a posteriori probability distribution. Thus, there is no correlation between $X$, $B_{1}$, and $B_{2}$ when conditioned on $U$, as required. \qed
%\end{appendices}
%\end{comment}

% Similar to the proof for the two-node case, using AFW inequality, the chain rule for entropy and the fact that conditioning decreases entropy for c-q states (in this case for the extended state  $\widehat{\rho}_{KX_kB_k}=\frac{1}{n}\sum_{k=1}^n\ketbra{k}\otimes \widehat{\rho}_{X_k,B_k,C_k}$) we can write:
% \begin{align}
% \abs{H\left(A^{n},B^{n},C^{n}\right)_{\widehat{\rho}}-\sum_{k=1}^{n}H\left(A_{k},B_{k}\right)_{\widehat{\rho}}} & \leq n\beta_n
% \label{Equation:AFW_Entropy_no_communication}
% \end{align}

% \begin{align}  I(A^n,B^n,C^n;U)=\sum_{k=1}^nI(A_k,)
% \end{align}
% According to the definition of 

% For the cardinality of $U$, based on the Caratheodory theorem \cite{csiszar2011information} we need $|\mathcal{U}|\leq [\mathrm{dim}(\mathcal{H}_A) \mathrm{dim}(\mathcal{H}_B) \mathrm{dim}(\mathcal{H}_C)]^2-1$ elements to preserve $H(ABC)$ and one additional element for preserving $H(ABC|U)$. 

\section{Summary and discussion}
We study coordination in three network models with classical communication links: 1) two-node network simulating  a c-q state, 2) no-communication network simulating a separable state, and 3)  a broadcast network simulating a c-q-q state.  Our models are motivated by quantum-enhanced Internet of Things (IoT) networks in which the communication links are classical \cite{notzel2020entanglement,9232550,burenkov2021practical,granelli2022novel}.
% We study coordination in three network models, two-node network simulating  a c-q state, no-communication network simulating a separable state, and  a broadcast network simulating c-q-q state. 
In our previous work \cite{NatorPereg:24c1}, we considered
entanglement coordination using quantum communication links, and our analysis therein was based on different tools compared to those used here.

Our findings generalize  classical results from \cite{cuff2008communication} and \cite{cuff2010coordination}, and also quantum results from \cite{state_generation_using_correlated_resource_2023}.
% The classical results corresponding to the discussed networks can be directly obtained by substituting the quantum systems by classical ones. For instance in the two-node case \cite{cuff2008communication}and the no-communication scenario \cite{cuff2010coordination}.
The no-communication and broadcast networks can easily be extended to $m$ encoders and decoders, respectively.
% The 3-node no-communication network and the 2 receivers broadcast network can be generalized to $m$ nodes in a straightforward way. For the no-communication network, for a desired state $\omega_{A_1\dots \A_m}$ we have:
% %
% \begin{align}
% \mathsf{C}_{\text{NC}}(\omega)=
% \inf_{\sigma_{UA_1\dots \A_m}\in\mathscr{S}_{NC}(\omega)}
% %\exists\rho_{XUB}\in \mathscr{S}\;s.t.\\
%  I(U;A_1\dots A_m)_\sigma\,.
% \end{align}
%
% where $\mathscr{S}_{NC}(\omega)$ is defined in a similar way to the 3-node no-communication case.  
% And for the broadcast network, with a desired state $\omega_{XB_1\dots B_2}$ we have
% %
% \begin{align}
% \mathcal{R}_{\text{BC}}(\omega)\triangleq
% \bigcup_{\sigma_{XUB_{1}\dots B_{m}}\in\mathscr{S}_{BC}(\omega)}
% \left\{ 
% \begin{array}{l}
% (R_0,R_1)\in\mathbb{R}^2:
% \\
% R_0  \geq I(X;U)_\sigma
% \\
% R_0+R_1  \geq
% \\{\;\;}I(XB_1\dots B_m;U)_\sigma
% \end{array}
% \right\} \,.
% \end{align} 
% %
% where $\mathscr{S}_{BC}(\omega)$ is defined in a similar way to the 2-receivers broadcast case.
%
The results are relevant for various applications, where the network nodes could represent classical-quantum sensors \cite{wang2022joint},  computers performing a joint computation task \cite{6G_quantum_2022,pereg2022classical}, 
or players in a nonlocal game \cite{Seshadri:23p,pereg2023multiple}.
Both this work and the previous work \cite{NatorPereg:24c1} can be viewed as a step forward in understanding coordination in networks that may comprise either classical or quantum resources. %communication links and correlation resources.

% Strong coordination in classical networks is important for cooperative games, as in theorem 7.1 in \cite{cuff2008communication}. Where in a game with two teams A and B, the first team consists of two players, and team B of one player. In team A, player~1 performs actions $x_i \in \mathcal{X}$ and player~2 performs actions $y_i \in \mathcal{Y}$. Team B performs actions $x_i \in \mathcal{Z}$. All the action spaces $\mathcal{X},\mathcal{Y}, \text{ and }\mathcal{Z}$ are finite. In each time step $i$, the payoff for team A is a described by a time invariant finite function $\Pi(x_i,y_i,z_i)$, this function is also the loss of team B. Each team wants to Player~1 can communicate with player~2 using a limited secure communication channel at rate $R$ by sending a message $U=u$ where $U$ satisfies the following Markov property $X^n-U-Y^n$ and $\abs{\mathcal{U}}\leq 2^{nR}$. Assume that team A wants to maximize the expected payoff for the worst case scenario for team B. Then the payoff for team A in the iteration $i$ is 
% %
% \begin{align}
%     \Theta_i \triangleq \min_{z\in \mathcal{Z}}\mathbb{E}\left[\Pi(x_i,y_i,z)|x^{i-1},y^{i-1}\right]
% \end{align}
% Cuff showed that for a given payoff value $\Theta$, the infimum over all achievable rates $R_1$ is 
% %
% \begin{align}
%   %R(\Theta)=  
% \end{align}

\section*{Acknowledgments}
  H. Nator and U. Pereg were supported by  Israel Science Foundation (ISF), Grants 939/23 and 2691/23, German-Israeli Project Cooperation (DIP), Grant
2032991, and  Nevet Program of the Helen Diller Quantum Center at the Technion, Grant 	2033613.
U. Pereg was also supported by the Israel VATAT Junior Faculty Program for Quantum Science and Technology through Grant 86636903, and the Chaya Career Advancement Chair, Grant 8776026.

%  \newpage
%  $\,$
% \newpage
{
% Align columns of bibliography
\balance 
\bibliography{references}
}
\end{document}